\documentclass[reprint,twocolumn,showpacs,aps,pre]{revtex4-1}
\usepackage{hyperref}
\usepackage[T1]{fontenc}
\usepackage{amssymb}
\usepackage{amsmath}
\usepackage{graphicx}
\usepackage{subfigure}
\newcommand{\bP}{\boldsymbol{\mathrm{P}}}
\newcommand{\bE}{\boldsymbol{\mathrm{E}}}
\newcommand{\bA}{\boldsymbol{\mathrm{A}}}
\newcommand{\bD}{\boldsymbol{\mathrm{D}}}
\newcommand{\bB}{\boldsymbol{\mathrm{B}}}

\newcommand{\bI}{\boldsymbol{\mathrm{I}}}
\newcommand{\bJ}{\boldsymbol{\mathrm{J}}}
\newcommand{\bZero}{\boldsymbol{0}}
\newcommand{\bp}{\boldsymbol{p}}
\newcommand{\by}{\boldsymbol{y}}
\newcommand{\bd}{\boldsymbol{d}}
\newcommand{\bv}{\boldsymbol{v}}
\newcommand{\bw}{\boldsymbol{w}}
\newcommand{\be}{\boldsymbol{e}}
\newcommand{\bz}{\boldsymbol{z}}
\newcommand{\bx}{\boldsymbol{x}}
\newcommand{\bu}{\boldsymbol{u}}
\newcommand{\bOne}{\boldsymbol{\mathrm{1}}}
\newcommand{\bxi}{\boldsymbol{\xi}}
\newcommand{\dmax}{d_{\max}}

\begin{document}
\title{The effect of network structure on phase transitions in queuing networks}
\date{\today}
\author{Norbert Barankai}
\author{Attila Fekete}
\author{G\'abor Vattay}
\affiliation{Department of the Physics of Complex Systems, E\" otv\" os
University, P\'azm\'any P\'eter S\'et\'any 1/A, H-1117 Budapest, Hungary}
\begin{abstract}
Recently, De Martino \emph{et al} \cite{DeMartinoEtal09a, DeMartinoEtal09b}
have presented a general framework for the study of transportation phenomena on complex
networks. One of their most significant achievements was a deeper understanding of the phase
transition from the uncongested to the congested phase at a critical traffic load.
In this paper, we also study phase transition in transportation networks using
a discrete time random walk model. Our aim is to establish a direct connection
between the structure of the graph and the value of the critical traffic load. Applying
spectral graph theory, we show that the original results of De Martino \emph{et
al} showing that the critical loading depends only on the degree sequence of
the graph---suggesting that different graphs with the same degree sequence have
the same critical loading if all other circumstances are fixed---is valid only
if the graph is dense enough. For sparse graphs, higher order corrections,
related to the local structure of the network, appear.
\end{abstract}
\maketitle
\section{Introduction}

During the past few decades, the physics community has witnessed enormous
progress in the research on complex networks \cite{AlbertBarabasi02, Newman03}.
Transport processes on networks represent an important class of dynamical
systems with a wide range of applications, including data traffic on the
Internet, vehicle traffic on highways, virus spread between hosts, or rumour
spread in social networks, to name but a few.  A simple, general model
can be used to describe these dynamical systems quite accurately,
in which particles are transported between the nodes of a network. 

In certain transport networks the particles are served by queues, residing at
the nodes of the network.  One of the most prominent examples is the Internet,
where the data packets play the role of the particles.  Queuing networks
exhibit several interesting phenomena, for example, below a critical traffic
intensity the system is in a free state and the average queue length fluctuates
around a finite value.  Above a critical traffic load, however, one or more
queues become congested and the average queue length diverges.

Various models have been developed to model the Internet traffic.
Deterministic and probabilistic routing strategies were compared in
\cite{OhiraSawatari98}.  The authors of \cite{ArenasDiazGuimera01} studied a
shortest-path routing model where the probability of packet transmission
depended on the queue lengths.  The authors of \cite{ZhaoLaiParkYe05} studied
how traffic congestion is affected by the capacity of the nodes in a simple
shortest path routing model.  A more elaborate routing strategy was studied in
\cite{Echeniqueetal05}, where the packets were forwarded to the neighbor that
minimized an effective distance to the packet's destination. Beyond that,
several attempts have been made to find routing strategies that are less
sensitive to congestion \cite{Melonietal10,Lingetal09,Lingetal10}. 

Packet level simulations of these models clearly indicate phase transitions
between the free and congested phases, and several characteristics of the
queuing networks exhibit power law dependence from the traffic intensity close
to the transition point \cite{OhiraSawatari98, ArenasDiazGuimera01,
ZhaoLaiParkYe05, Echeniqueetal05, Tretyakov98}.  The analytic description of
these models, however, is rather limited, because even the simplest routing
mechanism, namely the shortest-path routing, introduces non-local transport
dynamics to the system.

In recent years, extensive research has been undertaken to discover the
relationship between the topological properties of networks and the behavior of
the dynamical processes on them \cite{Strogatz01, DorogovtsevGoltsev08,Boccaletti06}.
A fundamental question is if a dynamical system shows phase transition
phenomena how does the structure of the network affect the phase transition.  In
models with non-local transport dynamics, however, the analytic description of the phase
transition has only been established for a few special networks, for example lattices
\cite{OhiraSawatari98,MukherjeeManna05} or Cayley trees
\cite{Tretyakov98,ArenasDiazGuimera01, ZhaoLaiParkYe05}.

In recent papers by De Martino \emph{et al} \cite{DeMartinoEtal09a,
DeMartinoEtal09b}, the authors studied the congestion phenomena in arbitrary
networks. The authors introduced a traffic aware congestion control mechanism
in their model, and modelled the transport process with a simple random walk
process, instead of a non-local routing mechanism.  It has been shown that the
model exhibits both first and second order phase transition, depending on the
parameters of a congestion control mechanism.

The main focus of our paper is to gain a deeper understanding of the
relationship between the structural properties of the underlying graph and the
congestion phenomena, that is the dynamics, and to give a generic
description of the phase transition point in an arbitrary network. 
Our model is similar to the one presented by De Martino \emph{et al}
\cite{DeMartinoEtal09a, DeMartinoEtal09b}.  We
approximate the particle transport process with a discrete time random walk
process, where packets are generated, absorbed and move randomly in the
network.  Moreover, we assume that the delivery of the particles is locally homogeneous,
that is the probability that a particle will be delivered from any node to its
neighbour is uniform, and we also assume that the queues are independent 
\cite{DeMartinoEtal09a}.

Our work differs from \cite{DeMartinoEtal09a, DeMartinoEtal09b} in two important
aspects, however.  Firstly, we do not consider any traffic control mechanism in our
study.  Secondly, instead of using time evolution equations, we apply a mean
field approximation in the long time limit and use spectral graph theory
\cite{Chung97} to connect the structure of the network with the traffic dynamics.
 
The paper is organized as follows.  After presenting our model in detail in
Sec.~\ref{sec:model}, we derive relationships that connect the critical
traffic loading with the parameters of the model (Sec.~\ref{sec:congestion}). 
In Sec.~\ref{sec:critical_load} and Sec.~\ref{sec:discussion}, we discuss the
 theoretical findings and we compare them to particle level
numerical simulations and numerical computations. Finally, we conclude our work in
Sec.~\ref{sec:conclusion}.  Some of the details of the analytic calculation are
presented in the Appendix.

\section{The model}
\label{sec:model}
\begin{figure}[t]
  \begin{center}
    \resizebox{.45\textwidth}{!}{\includegraphics{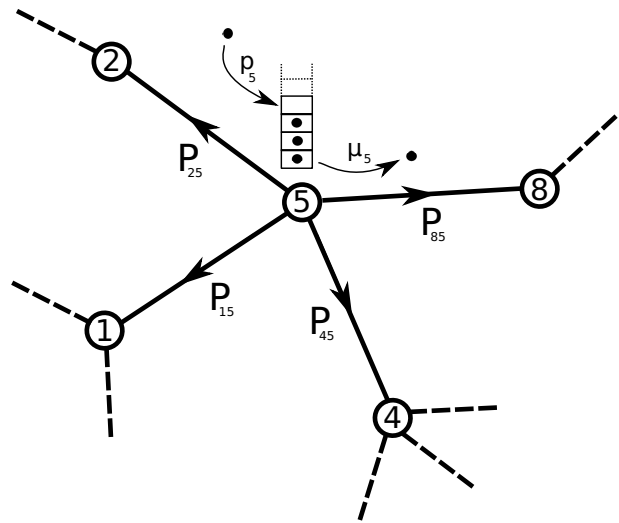}}
  \end{center}
  \caption{A schematic figure illustrating the dynamics of the model. Particles 
   are generated and absorbed at node $i$ by probability $p_i$ and $\mu_i$,
   respectively. The probability that a particle is delivered from node $i$ to 
   node $j$ is denoted by $P_{ji}$. (Color online.)} 
  \label{fig:network}
\end{figure}

We model the transportation network by a simple connected graph with $N$ nodes
and $M$ edges.  Moreover, the dynamics of the transport networks are modeled by
a discrete time stochastic process.  The rules of the stochastic process are
the following (see Fig.~\ref{fig:network}).  At each node of the graph, there is
a queue with infinite buffer capacity.  In each time step, the first particle
of each non-empty queue leaves the queue.  A particle at node $i$ will be
either absorbed (i.e.\ leaves the queuing network) with probability $\mu_i$, or
it is delivered to another queue at node $j$, adjacent to node $i$, with
probability $P_{ji}$. 

The probability that a particle will be absorbed at node $i$ can be expressed by
$\mu_i=1-\sum _j P_{ji}$. We will assume that the transition probability is 
constant in time. In locally homogeneous network dynamics, the transition
probabilities can be given by
\begin{equation}
  P_{ji} =\frac{1-\mu_i}{d_i},
\end{equation}
where $d_i$ is the degree of node $i$.

Note that as long as only the queue length statistics are concerned and not the
fate of individual particles (e.g.\ trajectories or travel times), the order in
which the particles leave the queues is irrelevant.  Therefore, individual
particles can be considered to be indistinguishable, and not only the first, but any
packet can be selected from the queues for delivery.

In each time step, after the delivery or absorption of the existing particles
in the system, new particles can also enter the queues randomly.  We assume
that the probability, $p_i$, that a new particle enters the system at node $i$
is also constant in time.  In addition, we will assume that the queuing
system is open, that is particles are generated and absorbed with non-zero
probability.

The queuing network can be in either a free or a congested state.  The network
is in free state if, after a transient period, the number of particles
in the system fluctuates around an average value. This stationary behavior
does not depend on the initial distribution of the length of the queues
\cite{Bolchetal06, Bhat08}.  In this case, the average number of particles
arriving to the system equals the average number of particles leaving the
network.  On the other hand, in the congested state the average number of
particles arriving to the system is greater than the average number of
particles absorbed. Therefore, in the congested state, the average number of
particles in the network will almost surely increase in time. This observation
suggests the definition of the order parameter
\begin{equation}
  \eta(p_1,\dots,p_N) =\lim_{t\to \infty} \frac{n(t+1)-n(t)}{\sum_i p_i},
\end{equation}
which measures the expected growth rate of the number of particles in the
system, $n(t)$, at time $t$, relative to the arrival rate of incoming particles,
$\sum_i p_i$ \cite{ArenasDiazGuimera01, GuimerGuilera02,
Huetal07, Zhaoetal05, Zhangetal07}. 

In the case of a stationary state, the order parameter is obviously zero,
whereas in the congested state it is greater than zero.  The
transition between the free and congested states can be characterized by the
critical probability, $\mathbf{p}_c=(p_1,p_2,\dots,p_N)^T$, where the expected
arrival rate of the incoming particles equals the expected rate of absorbed
particles at least at one of the queues.  It has been shown
earlier~\cite{Takayasu96,Takayasu00} that several characteristics of these
networks (e.g.~probability distribution of delay times, queue length
distribution, \emph{etc.}) show power law dependence on the loading
probability near the critical point, which suggests a close analogy with the
theory of phase transitions.

\section{Congestion in arbitrary networks}
\label{sec:congestion}
In this section we present an analytical estimation of the
critical point using a mean field approximation \cite{DeMartinoEtal09a}.

Let us suppose first that the queuing system is in equilibrium. In this
case, the expected number of particles arriving at each queue is
equal to the expected number of particles leaving the
queue, that is
\begin{equation}
  \xi_i=p_i+\Delta_i,
  \label{EQ:EQ1}
\end{equation}
where $\xi_i$ denotes the expected number of particles leaving the queue, 
$\Delta_i$ denotes the expected number of particles arriving to node $i$
from its neighbors in one time step, and $p_i$ is the arrival rate of the
particles at node $i$.  Since either zero or one particle leaves the queue in
each time step, $\xi_i$ is also equal to the probability that the
queue of node $i$ is not empty. Therefore, in the mean field approximation, 
where the queue lengths are independent, $\Delta_i$ can be calculated as
\begin{equation}
  \Delta_i=\sum_{j}P_{ij}\xi_j.
  \label{eq:mean_field}
\end{equation}

With standard vector and matrix notations we obtain that in the
uncongested, stationary phase the state vector $\bxi$ has to 
satisfy the equation $\left(\bE-\bP\right)\bxi=\bp$,
where $\bE$ denotes the identity matrix.
Since the matrix $\bE-\bP$ is invertible (see Appendix A), the loading
probabilities uniquely determine the components of the state vector $\bxi$
in the uncongested phase: 
\begin{equation}
  \bxi=(\bE-\bP)^{-1}\bp.
  \label{EQ:EQ2}
\end{equation}

The state of the network can be classified according to the state vector
$\bxi$. If the components of $\bxi$ satisfy the inequality
$\xi_i<1$ for all $i$, then the network is in an uncongested state, whereas  if 
there is at least one
node where $\xi_i=1$, the network is in the congested state.  Therefore, based on
this condition, the order parameter of the system can be calculated theoretically.

Note first that the balance equation~(\ref{EQ:EQ1}) cannot hold at the
congested nodes of the network, because the expected number of incoming
particles is greater than one, which is the maximum of the expected number of
outgoing particles at a queue.  Therefore, the congested queues grow steadily,
and these queues are never empty. It follows that $\xi_i=1$ for the congested queues,
and the expected growth rate of these queues can be given by $p_i+\Delta_i-1$.
Based on these observations we can develop an algorithm, presented in Appendix
\ref{APP:APP_C}, that can be used to calculat the order parameter
numerically for arbitrary networks and traffic load.  

In order to validate our model, we compared the order parameter obtained from
our algorithm with packet level simulations on the same network.  For the
comparison we used Barab\'asi--Albert (BA) \cite{BarabasiAlbert99},
Erd\H{o}s--R\'enyi (ER) \cite{ErdosRenyi60} and Watts--Strogatz (WS)
\cite{WattsStrogatz98} networks.  The loading probability was the same at every
node of the network, and both the absorption probabilities and the elements of
the transition matrix $\bP$ were random numbers distributed uniformly between
zero and one. 

\begin{figure}[t]
  \subfigure[BA network ($m=2$)]{
    \includegraphics[width=.4\textwidth]{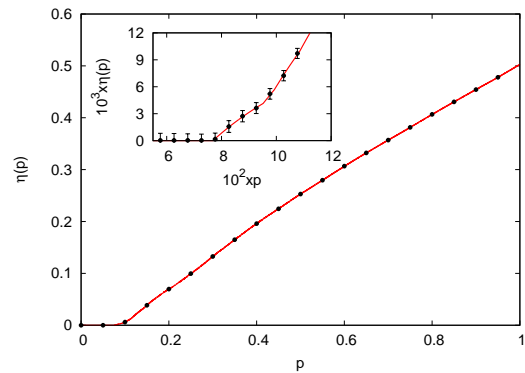}
  }
  \subfigure[ER network ($p_{\mathrm{ER}}=0.4$)]{
    \includegraphics[width=.4\textwidth]{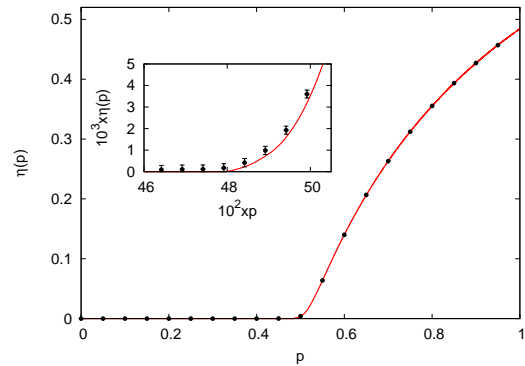}
  }
  \subfigure[WS network ($z=16$ and $q_{WS}=0.1024$)]{
    \includegraphics[width=.4\textwidth]{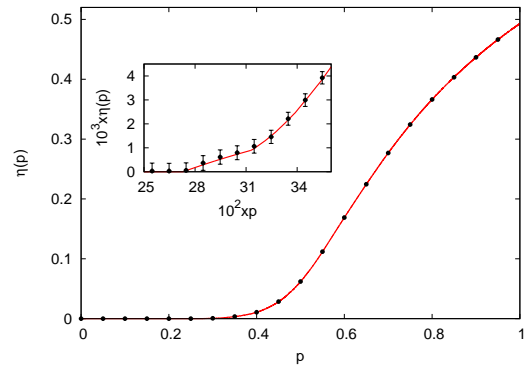}
  }
  \caption{The order parameter as a function of homogeneous loading
    probabilities. The graphs show numerical calculations (solid line)
    and particle based simulations (black points) on three distinct
    networks. All graphs had $N=500$ nodes. 
    (Color online.)}
 \label{FIG:FIG2}
\end{figure}

The results are shown in Fig.~\ref{FIG:FIG2}. It can be seen that the 
theoretical curve, computed by our numerical method, fits very well to the
values of the order parameter determined by simulations.

\begin{figure}[t]
  \includegraphics[width=.4\textwidth]{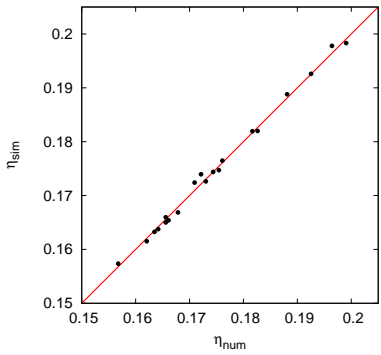}
  \caption{Comparison of the order parameter computed numerically with our
    algorithm and obtained from simulation with inhomogeneous loading
    probabilities on an ER graph($N=500$ and $p_{ER}=0.4$). (Color online.)}
  \label{FIG:FIG3}
\end{figure}

We also validated our results for inhomogeneous traffic loads. For this purpose
we generated several realizations of random loading vectors, $\bp$, and
transition matrix $\bP$ on the same ER graph and compared the order parameter
calculated by simulations and numerical computations. Results are shown in
Fig.~\ref{FIG:FIG3}.  The simulations agree with our numerical method very well.

\section{Critical traffic load in arbitrary networks}
\label{sec:critical_load}

The main difficulty of using Eq.~(\ref{EQ:EQ2}) is the computational
complexity of inverting the matrix $\bE-\bP$.  Moreover, even if the matrix can
be inverted numerically for a particular network, the dependence of the
critical point on the network structure and the traffic load remains obscure. In
the case of large irregular networks approximations are needed to describe the
phase transition analytically. 

For a detailed analysis of topological effects on the critical load $\bp_c$, let us consider
networks with random-walk-like particle transport with homogeneous absorption
probabilities. In this case, if nodes $i$ and $j$ are connected in the network, the
transition probability from node $i$ to node $j$ is $P_{ji} =(1-\mu)/d_i$, where $d_i$ is
the degree of node $i$, and $\mu$ is the absorption probability. Using the
adjacency and degree matrix of the network, $\bP$ can be written in a very
compact form: 
\begin{equation}
  \bP=(1-\mu)\bA\bD^{-1}. 
\end{equation}

Using the spectral decomposition of the transition matrix, we can write
\begin{equation}
  \bA\bD^{-1}=\sum_{k=1}^N\lambda_k\bv_k\bw^T_k.
\end{equation}
where $\bw_k$ and $\bv_k$ are the left and right eigenvectors of 
$\bA\bD^{-1}$, respectively, and $\lambda_k$ is the corresponding eigenvalue. Note that
the symmetric matrix $\bD^{-1/2}\bA\bD^{-1/2}$ has the same eigenvalue 
spectrum as the transition matrix $\bA\bD^{-1}$. 
Indeed, if $\boldsymbol{u}_k$ is an
eigenvector of the later matrix with eigenvalue $\lambda_k$, then
$\bv_k=\bD^{1/2}\boldsymbol{u}_k$ will be the right, and
$\bw_k=\bD^{-1/2}\boldsymbol{u}_k$ will be the left eigenvector of
$\bA\bD^{-1}$ with the same eigenvalue \cite{Lovasz93}.

The transition matrix $\bA\bD^{-1}$ is well known in the theory of random walks
on graphs \cite{Lovasz93}.  Its eigenvalues satisfy the inequality $-1\le
\lambda_k \leq 1$ \cite{Lovasz93}.  The largest eigenvalue is always
$\lambda_1=1$ and it is degenerate only if the graph is not connected.
Moreover, the smallest eigenvalue is equal to $-1$ if and only if the graph is
bipartite \cite{Lovasz93}. In the following we will assume that the graphs
under study are connected. 

It is easy to see that the normalized eigenvector of $\bD^{-1/2}\bA\bD^{-1/2}$ 
corresponding to $\lambda_1=1$ is
$\bu_1=\bD^{1/2}\bOne/\sqrt{2M}$, where $M$ is
the number of links in the network, and $\bOne=(1,1,\dots,1)^T$.
It follows that the left and right eigenvectors of $\bA\bD^{-1}$ corresponding to
$\lambda_1=1$ are
$\bw_1=\bOne^T/\sqrt{2M}$ and $\bv_1=\bd/\sqrt{2M}=\bD\bOne/\sqrt{2M}$, respectively.  

Using the power series expansion of the function $1/(1-x)$ and the spectral
decomposition of the transition matrix, we can formally calculate the inverse
of $\bE-\bP$:
\begin{equation}
  (\bE-\bP)^{-1}=\sum_{k=1}^N\frac{1}{1-(1-\mu)\lambda_k}\bv_k\bw^T_k.
\end{equation}

The above sum can be split into three parts that have behave differently 
as $\mu$ is varies:
\begin{equation}
  \sum_{\lambda_k=0}\bv_k\bw^T_k
  +\frac{1}{\mu}\frac{\bd\bOne^T}{2M}
  +\sum_{\lambda_k\neq 0,1}\frac{1}{1-(1-\mu)\lambda_k}\bv_k\bw^T_k.
\label{3T}
\end{equation}
The first term, corresponding to the eigenvalues $\lambda_i=0$, is independent
of $\mu$. The second term, which belongs $\lambda_1=1$, becomes singular as
$\mu\rightarrow 0$.  The last term, which consists of all the eigenvalues that
are neither zero nor one, is finite for every values of $\mu$. 

\section{Discussion}
\label{sec:discussion}

\subsection{Low absorption levels}

In the case of low absorption levels, the second term dominates in 
(\ref{3T}). Therefore, we obtain
\begin{equation}
  \bxi=\left(\bE-\bP\right)^{-1}\bp\simeq\frac{\bd}{\overline{d}}\frac{\overline{p}}{\mu},
  \label{EQ:EQ3}
\end{equation}
where $\overline{p}=\sum_ip_i/N$ and $\overline{d}=2M/N$. Moreover, if the
loading is homogeneous, i.e.\ $\bp=p\bOne$, the critical loading probability is
\begin{equation}
  p_c=\frac{\overline{d}}{\dmax}\mu,
  \label{EQ:EQ4}
\end{equation}
where $\dmax$ is the maximal degree in the graph.  

Note that in the low absorption limit the critical point depends only on the
relative spread of the degree sequence, i.e.\ $\dmax/\overline{d}$, and not on
the absolute scale of the degrees.  In particular, in the case of regular
graphs, where each node has the same degree, the relative spread is
$\dmax/\overline{d}=1$, so the critical point, $p_c=\mu$, is independent of the
degree of a regular graph.  Furthermore, $\dmax$ is always greater than or
equal to $\overline{d}$, and equality holds iff the graph is regular.
Therefore, the critical point $p_c$ is the highest in regular graphs at a
given absorption level. 

\subsection{High absorption levels}
If the absorption probability $\mu$ is close to one, then $(1-\mu)\lambda_k$ is
close to zero, so the denominators of the third term in (\ref{3T}) can be
approximated by one.  Since the eigensystem of $\bA\bD^{-1}$ is complete, we
obtain that
\begin{equation}
  (\bE-\bP)^{-1}\simeq\bE+\frac{1-\mu}{\mu}\frac{\bd\bOne^T}{2M},
\end{equation}
and the state vector can be approximated by
\begin{equation}
  \bxi\simeq\bp+(1-\mu)\frac{\bd}{\overline{d}}\frac{\overline{p}}{\mu}.
  \label{EQ:EQ5}
\end{equation}
Note that we kept the $1/\mu$ factor in (\ref{EQ:EQ5}). This way
Eq.~(\ref{EQ:EQ5}) can reproduce Eq.~(\ref{EQ:EQ3}) in the small absorption limit,
since $\bp$ can be neglected compared to the second term, which diverges if
$\mu\rightarrow 0$.

In the case of homogeneous loading the critical loading probability is
\begin{equation}
p_c\simeq \frac{\mu}{\mu+(1-\mu)\dmax/\overline{d}},
\label{EQ:EQ6}
\end{equation}
which depends again only on $\dmax/\overline{d}$, the relative spread of the
degree sequence. It is remarkable that the critical point is almost completely
independent of the fine details of the network structure in both the small and
high absorption limit.  

\subsection{Intermediate absorption levels}

Numerical simulations presented later in this section show that
Eq.~(\ref{EQ:EQ5}) is valid not only for small and large values of $\mu$, but
also for intermediate values if the edge density of the graph is large. The
reasons are the following. The approximation that leads to Eq.~(\ref{EQ:EQ5}) is
the assumption that the term $(1-\mu)\lambda_k$ is close to zero for those
$\lambda_k$, that are neither equal to zero nor one. This approximation is
valid not only when $\mu$ is close to one, but also if the second largest
absolute eigenvalue is close to zero. 

If the graph is non-bipartite, the second largest absolute eigenvalue is
related to $\tau$, the characteristic time until a particle reaches the
stationary distribution of a random walk, as
$1/\tau=\max\{|\lambda_2|,|\lambda_N|\}$.  Numerical calculations, presented in
Section~\ref{subsec:Error_estimation}, show that $1/\tau$ decreases when the
average degree increases and remains constant with small fluctuations if the
average degree is fixed.  Therefore, in the case of dense networks, it is plausible
to use the lowest terms of the power series expansion
\begin{equation}
\frac{1}{1-(1-\mu)\lambda_k}=1+(1-\mu)\lambda_k+(1-\mu)^2\lambda_k^2+\dots.
\label{PE}
\end{equation}
in the third term of Eq.~(\ref{3T}).  The zeroth order term of the series
expansion reproduces Eq.~(\ref{EQ:EQ5}).  If the network is not dense, like
many real networks, or the absorption level is in the intermediate range we
need to consider the first order term in the power series expansion
(\ref{PE}) as well.  Since bipartite and non-bipartite graphs are qualitatively
different, we will discuss the two cases separately. 

\subsubsection{Non-bipartite graphs}
It is easy to see that the first order correction to
$\left(\bE-\bP\right)^{-1}$ is $(1-\mu)(\bA\bD^{-1}-\bd\bOne^T/2M)$. Using this
correction we obtain that the state vector can approximated by 
\begin{equation}
  \bxi\simeq \bp+(1-\mu)\bA\bD^{-1}\bp+(1-\mu)^2\frac{\bd}{\overline{d}}\frac{\overline{p}}{\mu}.
\end{equation}

\begin{figure*}[t]
  \subfigure[The critical load, $p_c$, in a BA network.]{
    \includegraphics[width=.45\textwidth]{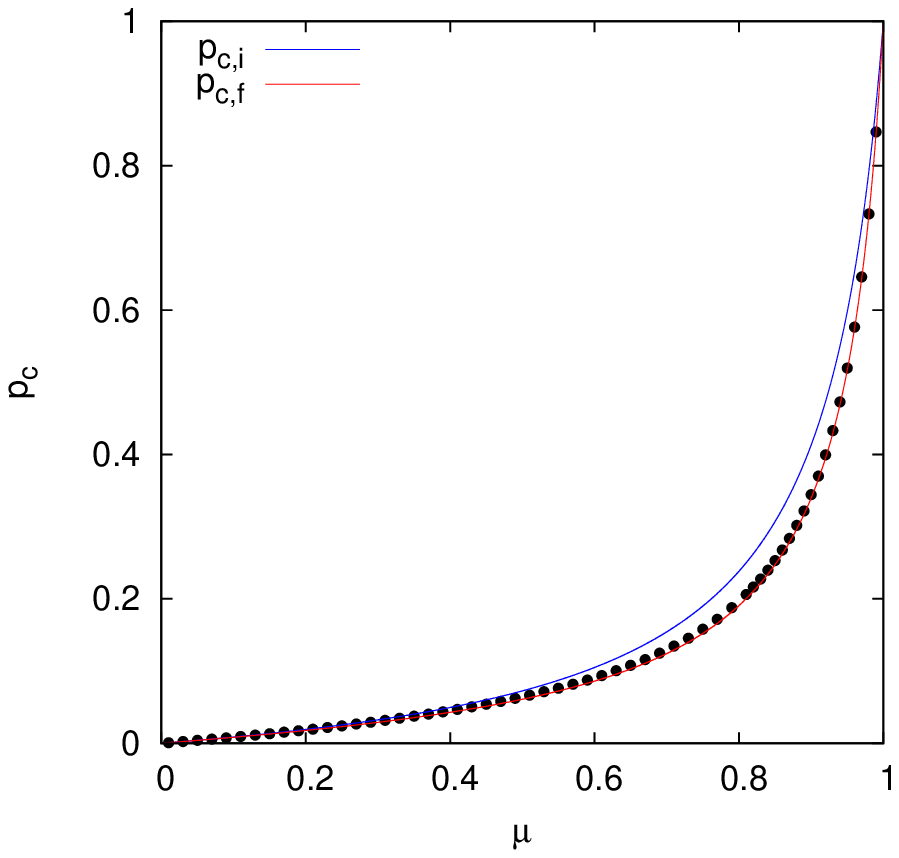}
}
  \subfigure[The critical load, $p_c$, in an ER network.]{
    \includegraphics[width=.45\textwidth]{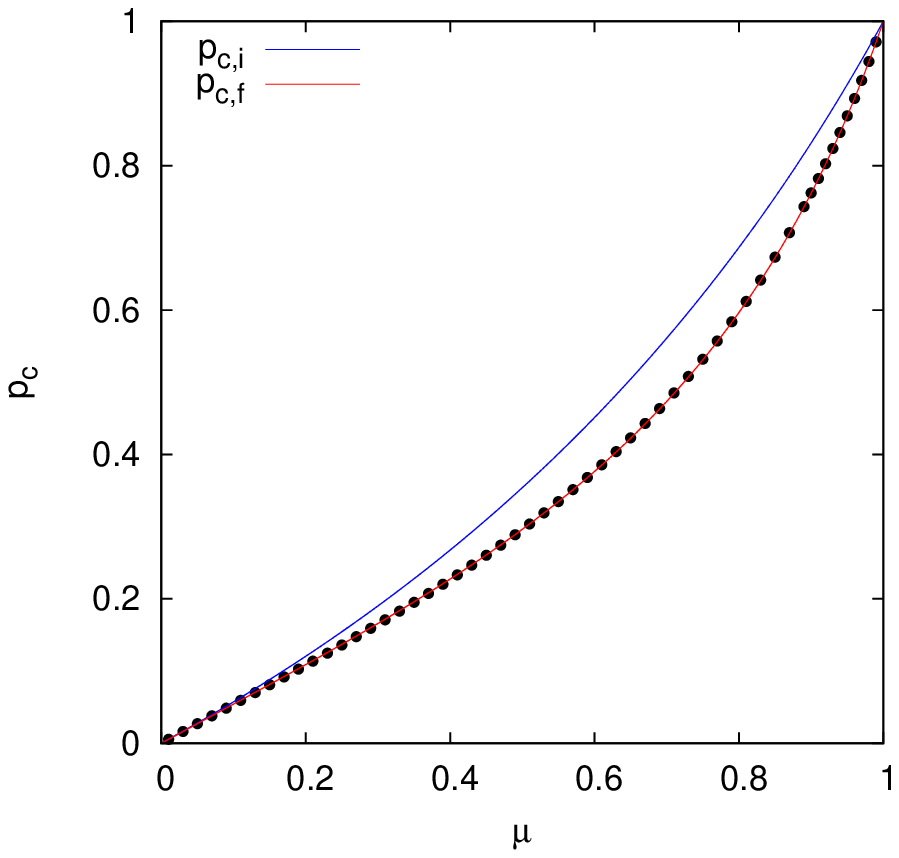}
}
  \subfigure[The absolute error, $\Delta p_c$ in a BA network.]{
    \includegraphics[width=.45\textwidth]{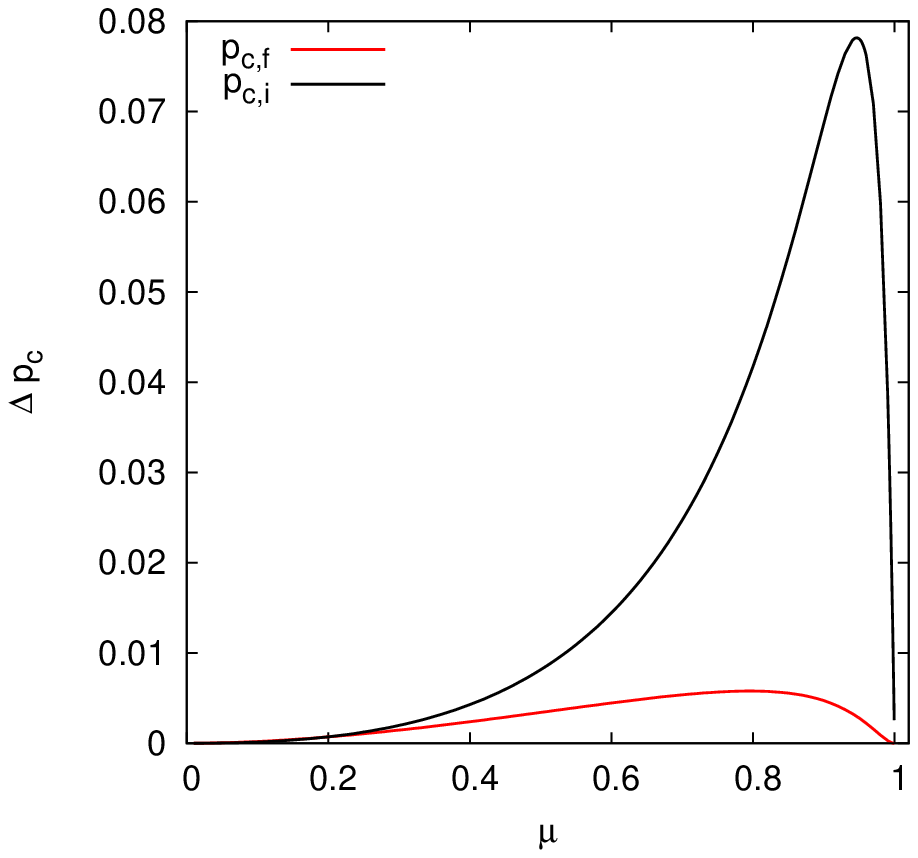}
}
  \subfigure[The absolute error, $\Delta p_c$ in an ER network.]{
    \includegraphics[width=.45\textwidth]{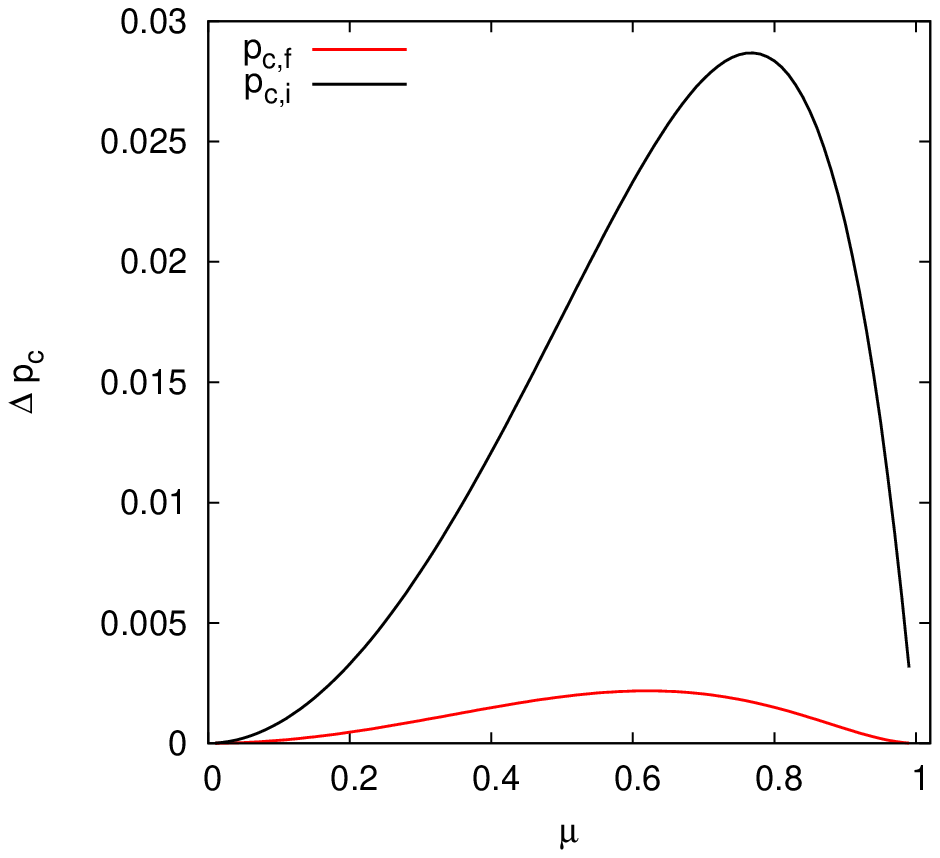}
}
  \caption{Numerical validation of the analytic results in BA ($N=512$ nodes,
$m=2$) and ER ($N=64$ nodes, $p=0.1$) networks.  The continuous lines $p_{c,i}$
and $p_{c,f}$ represent Eqs.~(\ref{EQ:EQ6}) and (\ref{EQ:EQ7}), respectively.
Numerical data, obtained by solving Eq.~(\ref{EQ:EQ2}) numerically, are
represented by dots. (Color online.)}
  \label{FIG:FIG4}
\end{figure*}

In order to study the effects of the topology on the critical traffic load, let
us consider homogeneous loading probabilities. After straightforward
calculations we obtain that in this case the $i$th component of the
state vector is
\begin{equation}
  \frac{\xi_i}{p}\simeq 1+\frac{(1-\mu)^2}{\mu}\frac{d_i}{\overline{d}}+(1-\mu)\frac{d_i}{h_i}.
\end{equation}
where $h_i$ denotes the harmonic mean of the degree of the neighbors of the $i$th node,
\begin{equation}
h_i=\left(\frac{1}{d_i}\sum_{j\in\mathcal{N}(i)}\frac{1}{d_j}\right)^{-1},
\end{equation}
and $\mathcal{N}(i)$ is the set of neighbors of node $i$. Consequently, we obtain
that the critical loading probability is
\begin{equation}
  \frac{1}{p_c}\simeq 1+\underset{i}{\mathrm{max}}\left\{\frac{(1-\mu)^2}{\mu}
  \frac{d_i}{\overline{d}}+(1-\mu)\frac{d_i}{h_i}\right \},
\label{EQ:EQ7}
\end{equation}
which is one of the main results of our paper.

The main novelty of Eq.~(\ref{EQ:EQ7}) is that it shows that the critical point
of the phase transition is determined not only by the spread of the degree
sequence, $\dmax/\overline{d}$, like in the low and high absorption limit,
but also by $h_i$, which depends on the local structure of the network. 

We validated our result on BA and ER networks. We calculated the critical load 
$p_c$ by inverting Eq.~(\ref{EQ:EQ2}) numerically and then compared the
result with Eq.~(\ref{EQ:EQ6}) and Eq.~(\ref{EQ:EQ7}) on the same graph
with the same absorption level. 

The results are presented in Fig.~\ref{FIG:FIG4}.  It can be seen that the
zeroth order approximation (\ref{EQ:EQ6}) is valid only for small and large
absorption levels. On the contrary, the first order approximation
(\ref{EQ:EQ7}) fits the numerical data closely on the whole range of absorption
levels.  In order to emphasize the clear advantage of Eq.~(\ref{EQ:EQ7}), the
absolute difference between formulas (\ref{EQ:EQ6}) and (\ref{EQ:EQ7}) and the
numerical data is also shown in Fig.~\ref{FIG:FIG4}.


\subsubsection{Bipartite graphs}
The nodes of a bipartite graph can be divided into two disjoint groups,
$\mathcal{G}_1$ and $\mathcal{G}_2$, in such a way that the edges of the graph
connect only nodes from different groups. Since trees are bipartite
graphs, bipartite graphs are extremely important from the practical point of view. 

Suppose that there are $N_1$ nodes in $\mathcal{G}_1$ and $N_2$ nodes in
$\mathcal{G}_2$, and denote the average degree and loading probability in
$\mathcal{G}_i$ ($i=1,2$) are $\overline{d}_i=M/N_i$ and $\overline{p}_i=\sum_{k\in
\mathcal{G}_i}p_k/N$, respectively. 

The critical traffic load in bipartite graphs can be calculated similarly to
the non-bipartite graphs.  The calculation is based on the symmetry of the
spectra of the transition matrix.  Here, we only summarize the results for
homogeneous loading, details of the calculation and the case of heterogeneous
loading are presented in Appendix \ref{APP:APP_B}. 


In the case of a bipartite graph we obtain that each partition defines a separate
critical loading probability.  For low or high absorption levels and
homogeneous traffic load we obtain, analogously to Eq.~(\ref{EQ:EQ6}), that
\begin{equation}
\begin{aligned}
\frac{1}{p_c^{(1)}}\simeq &1+\frac{1}{\mu}\frac{(1-\mu)^2/\overline{d}_1+(1-\mu)/\overline{d}_2}{2-\mu}\dmax^{(1)},\\
\frac{1}{p_c^{(2)}}\simeq &1+\frac{1}{\mu}\frac{(1-\mu)^2/\overline{d}_2+(1-\mu)/\overline{d}_1}{2-\mu}\dmax^{(2)},
\end{aligned}
\label{EQ:EQ8}
\end{equation}
where $\dmax^{(i)}$ is the maximal degree in $\mathcal{G}_i$. The critical
loading  probability of the whole network is the lesser of the two:
$p_c=\min(p_c^{(1)}, p_c^{(2)})$ . 

This result is very similar to the case of non-bipartite graphs. It can be
seen that $p_c$ depends only on the absorption level, $\mu$, and global
properties of the graph, namely the mean $\overline{d}_i$ and the maximal
degree $\dmax^{(i)}$, which can be obtained from the degree
sequences of $\mathcal{G}_1$ and $\mathcal{G}_2$ straightforwardly.

In the case of intermediate absorption levels, the zeroth order approximation of 
critical traffic load, presented in (\ref{EQ:EQ8}), has to be corrected. 
Similarly to non-bipartite graphs, the first order
correction, obtained from the spectral decomposition of the transition matrix,
is proportional to $d_i/h_i$. The critical loading probabilities $p_c^{(1)}$ 
and $p_c^{(2)}$, including the first order corrections and corresponding 
to the two sub-components of a bipartite graph, are
\begin{equation}
\begin{aligned}
\frac{1}{p_c^{(1)}}\simeq 1+\underset{i\in\mathcal{G}_1}{\mathrm{max}}
\Bigg\{&\frac{1}{\mu}\frac{(1-\mu)^2/\overline{d}_1+(1-\mu)^3/\overline{d}_2}{2-\mu}d_i+\\
&+(1-\mu)d_i/h_i\Bigg\},\\
\frac{1}{p_c^{(2)}}\simeq 1+\underset{i\in\mathcal{G}_2}{\mathrm{max}}
\Bigg\{&\frac{1}{\mu}\frac{(1-\mu)^2/\overline{d}_2+(1-\mu)^3/\overline{d}_1}{2-\mu}d_i+\\
&+(1-\mu)d_i/h_i\Bigg\}.
\end{aligned}
\label{EQ:EQ9}
\end{equation}
Consequently, the critical loading probability of the whole network is 
$p_c=\min\left(p_c^{(1)}, p_c^{(2)}\right)$.

\begin{figure}[t]
  \subfigure[The critical load, $p_c$, as a fuction of 
  the absorption level, $\mu$.]{
    \includegraphics[width=.45\textwidth]{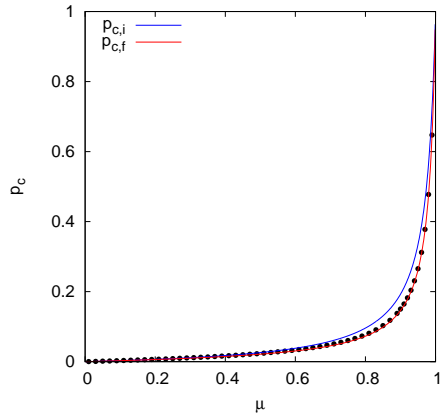}
}
  \subfigure[The absolute error of the critical load, 
  $\Delta p_c$, as a function of the absorption level, $\mu$.]{
    \includegraphics[width=.45\textwidth]{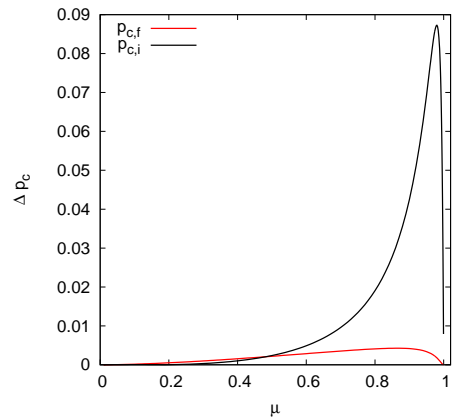}
}
  \caption{Numerical validation of the analytic results on a BA tree 
($N=1024$ nodes, $m=1$).  The continuous lines $p_{c,i}$
and $p_{c,f}$ represent Eqs.~(\ref{EQ:EQ8}) and (\ref{EQ:EQ9}), respectively.
Numerical data, obtained by solving Eq.~(\ref{EQ:EQ2}) numerically, are
represented by dots. (Color online.)}
  \label{FIG:FIG5}
\end{figure}

In order to validate our results, we calculated the critical load $p_c$ by
inverting (\ref{EQ:EQ2}) numerically.  For validation, we used BA scale-free
trees, grown by preferential attachment, and compared the numerical data with
(\ref{EQ:EQ8}) and (\ref{EQ:EQ9}) with the same graph and absorption level.
The results are shown in Fig.~\ref{FIG:FIG5}.  Eq.~(\ref{EQ:EQ9}), which
includes first order corrections, fits the numerical data closely not only at
low and high absorption levels, like Eq.~(\ref{EQ:EQ8}), but also in the
intermediate absorption range. 

\subsection{Error estimation in the large graph limit}
\label{subsec:Error_estimation}

We have seen that in the case of non-bipartite graphs the largest absolute
eigenvalue is $|\lambda_1|=1$, whereas in case of bipartite graphs it is
$|\lambda_1|=|\lambda_N|=1$.  In the power series expansion (\ref{PE}) we
considered the largest absolute eigenvalue precisely, and neglected the higher
order terms for $|\lambda|<1$.  In this section we discuss the validity
of our approximation, and show numerically how the precision of our model
depends on the graph properties. 

The error of the derived formulas depends on the magnitude of the higher order
terms that we neglected in the power series expansion (\ref{PE}). These terms
can be bounded from above with the second largest absolute eigenvalue, that is
$\max\{|\lambda_2|, |\lambda_N|\}$ for non-bipartite graphs and $\lambda_2$ for
bipartite graphs.  The smaller the second largest absolute eigenvalues are, the
smaller the error of Eq.~(\ref{EQ:EQ7}) and Eq.~(\ref{EQ:EQ9}) is.

Although it is easy to manipulate the eigenvalues $\lambda_k$ formally, it is
difficult to see how the eigenvalues depend on the graph properties.  In order
to see more easily how the error depends on the graph properties, let us
introduce the mixing time from the theory of random walk on graphs.

The mixing time, $\tau$, is the expected time until a particle, performing
random walk on a graph, reaches a stationary distribution.  It can be shown that
the mixing rate, the reciprocal of the mixing time, is precisely equal to the
second largest eigenvalue, i.e.  $1/\tau=\max\{|\lambda_2|, |\lambda_N|\}$ for
non-bipartite graphs and $1/\tau=\lambda_2$ for bipartite graphs
\cite{Lovasz93}. Therefore, it is plausible to conclude that for graphs which
have small mixing rate, the error will be also small.


There is no known formula in the literature on how the mixing rate depends on
the graph parameters in general \cite{Lovasz12}.  However, our numerical
experiments showed that the mixing rate depends strongly on the edge density,
$M/N=\overline{d}/2$.  In particular the mixing rate decreases as the edge
density increases.  In Fig.~\ref{FIG:FIG7} we can see the relative error of the
derived formulas, $\Delta p_c/p_c$, as the function of various graph
parameters.  In the inset we can see the mixing rate as the function of the
corresponding graph parameter.

In the case of a BA network, for example, the edge density is $M/N\simeq m$,
where $m$ is the number of edges connecting the new nodes to the graph in
preferential attachment. This means that the mixing rate, $1/\tau$, remains
constant with small fluctuations if $m$ is fixed, even if $N\to\infty$. This
phenomenon is the same in bipartite and non-bipartite graphs.  For example, one
obtains a BA scale-free tree, which is a bipartite graph, if $m\equiv1$, In
this case the error of the derived formulas is also constant, and it will not
decrease even in the thermodynamic limit.  

\begin{figure}[t]
  \subfigure[BA tree with $m=1$ fixed and $N$ varied.]{
  \includegraphics[width=.4\textwidth]{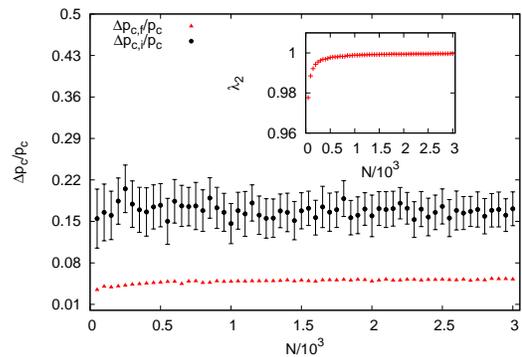}
  \label{FIG:FIG6a}
}
  \subfigure[BA network with $m=2$ fixed and $N$ varied.]{
  \includegraphics[width=.4\textwidth]{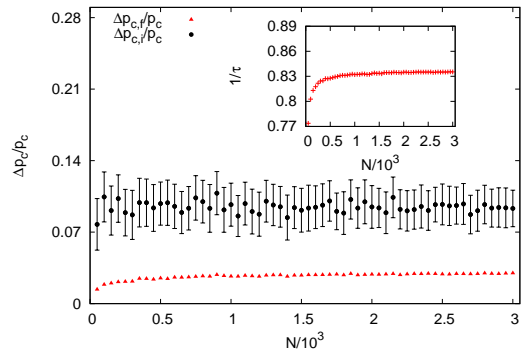}
  \label{FIG:FIG6b}
}
  \subfigure[BA network with $N=1024$ fixed and $m$ varied.]{
  \includegraphics[width=.4\textwidth]{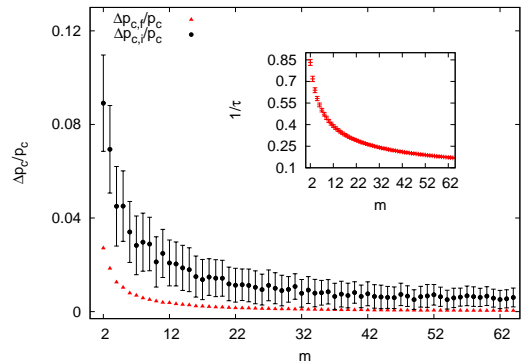}
  \label{FIG:FIG6c}
}
  \caption{The relative error, $\Delta p_c/p_c$, of the derived formulas as a
function of $N$ and $m$ in BA networks with absorption rate $\mu=0.8$. The
black dots and red triangles show the error of the derived zeroth and first
order approximations, respectively. The insets show the dependence of the
mixing rate, $1/\tau$ on $N$ and $m$.  Data points were averaged over $32$
graph realization in each case. (Color online.)}
  \label{FIG:FIG6}
\end{figure}

\begin{figure}[t]
  \subfigure[ER network with $N=1024$ fixed and $p_{ER}$ varied.]{
  \includegraphics[width=.4\textwidth]{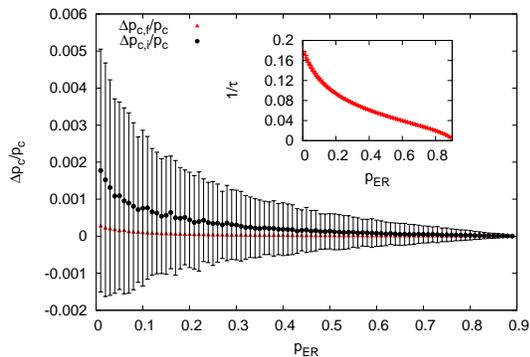}
  \label{FIG:FIG7a}
}
  \subfigure[ER network with $p_{ER}=0.5$ fixed and $N$ varied.]{
  \includegraphics[width=.4\textwidth]{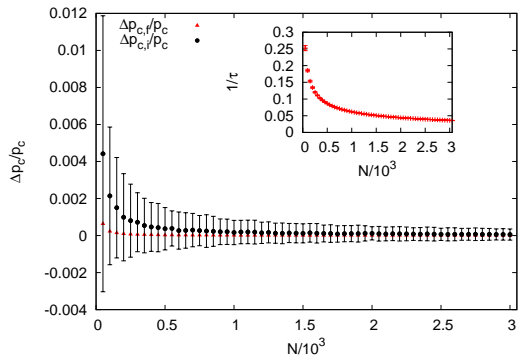}
  \label{FIG:FIG7b}
}
  \subfigure[ER with $Np_{ER}=49$ fixed and $N$ varied.]{
  \includegraphics[width=.4\textwidth]{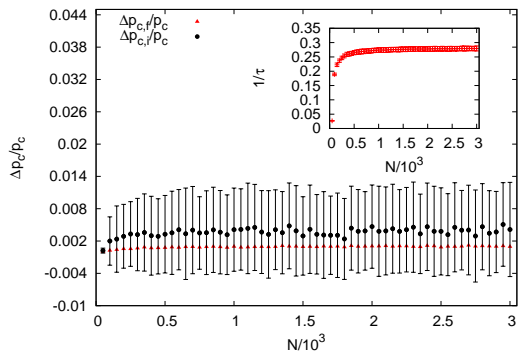}
  \label{FIG:FIG7c}
}
  \caption{The relative error, $\Delta p_c/p_c$, of the derived formulas as a
function of $N$ and $p_{ER}$ in ER networks with absorption rate $\mu=0.8$
(Figs.~\ref{FIG:FIG7a} and \ref{FIG:FIG7b}) and $\mu=0.5$
(Fig.~\ref{FIG:FIG7c}).  The black points and red triangles show the error of
the derived zeroth and first order approximation, respectively. The insets show
the dependence of the mixing rate, $1/\tau$, on $N$ and $p_{ER}$.  Data points
were averaged over $32$ graph realizations in each case. (Color online.)}
  \label{FIG:FIG7}
\end{figure}

In Fig.~\ref{FIG:FIG6a} and Fig.~\ref{FIG:FIG6b} we can see BA networks with
$m=1$ and $m=2$ fixed, and $N$ varied.  These cases correspond to a bipartite
and a non-bipartite graph, respectively. We can see that in both cases both the
relative error and the mixing rate tends to a fixed value as $N\to\infty$.  On
the other hand, in Fig.~\ref{FIG:FIG6c} the size of the graph is fixed, and $m$
is varied.  We can see that the mixing rate increases as $m$ increases and,
at the same time, the relative tends to zero.

The construction of ER networks is fundamentally different from the BA graphs.
In the case of ER networks, the edge density is $M/N=p(N-1)/2\simeq pN/2$.
Therefore, the edge density increases, and the relative error tends to zero, if
either $p$ or $N$ is increased, and the other parameter is fixed.  Note that in
case of $p=1$ the ER network is a complete graph, in which case the derived
formulas are exact.  In contrast, the relative error of the derived formulas
tends to a fixed value if $pN$ is fixed.  

Numerical simulations carried out on ER networks are shown in Fig.~\ref{FIG:FIG7}.
We can see that the simulation results confirm our assumption that the relative error 
of the derived formulas decreases if the density of the network increases, and
remain fixed if the density is fixed.

\section{Conclusion}
\label{sec:conclusion}
In this paper, we studied congestion phenomena in queuing networks.  We
analyzed how the critical point of the phase transition between free and
congested phases is influenced by the topological properties of the network.
In order to study the influence of the network structure on the traffic
dynamics in an arbitrary network, we neglected congestion control mechanisms,
and we modeled the particle transport by a simple Markovian random walk. 

In our model the critical traffic load in the network can be controlled by the
absorption level of the particles.  We first derived Eq.~(\ref{EQ:EQ6}), the
zeroth order approximation for the critical traffic load for low and high
absorption levels.  This result has been obtained by De Martino \textit{et al}
\cite{DeMartinoEtal09a} on a model with a congestion control mechanism
included. Our result confirms their finding that at the critical point the
details of the congestion control mechanism are less important in some cases.

In our paper we also showed that in the case of intermediate absorption levels
the zeroth order formula is not valid, and higher order corrections are needed.
We derived Eq.~(\ref{EQ:EQ7}) which incorporates the first order corrections to
Eq.~(\ref{EQ:EQ6}) and improves the precision of the critical point
considerably. In contrast to the zeroth order formula, higher order corrections
include not only the global properties of the degree sequence, i.e. the mean
and maximum degree, but also the local information on the network structure.
The improvement achieved by the higher order correction was validated by
numerical simulations. 

We also demonstrated that in the case of intermediate absorption levels the
structure of the network can have dramatic effects on the analytic behavior of the
critical point. We showed, in particular, that one must pay special attention 
when considering a bipartite graph, because the spectra of bipartite graphs
are symmetrical. We derived Eq.~(\ref{EQ:EQ9}) and showed that the critical
point in a bipartite graph is the maximum of the critical points of its
sub-components. 

Finally, we investigated the validity of our model.  We showed that the higher
order terms that were neglected during our calculations depend on the spectral
gap, which can also be expressed by the mixing rate in the graph.  We presented
numerical arguments that the mixing ratio, that is the precision of our
approximations, strongly depends on the edge density $M/N$. This empirical fact
is well known in the mathematical community, but up to now, as far as we know,
there is no rigorous proof of the phenomenon. We would like to examine this
question in detail in our future work.

\section{Acknowledgements}

This work was partially supported by the National Science Foundation OTKA 7779, the
National Development Agency (TAMOP 4.2.1/B-09/1/KMR-2010-0003) and the EU FIRE
NOVI project (Grant No. 257867).  We are also grateful to professor L\'aszl\'o
Lov\'asz for the useful discussions on the mixing rate of graphs. 

\appendix
\section{}
\label{APP:APP_A}
Let us suppose that $\bE-\bP$ is not invertible, which is equivalent to the
statement that $\bP$ has a normalized eigenvector $\boldsymbol{x}$
with eigenvalue $1$. In this case, using the assumption that there is
at least one node where $\sum_{i}P_{ij}<1$ (e.i.~there is absorption
at least at one node), the following inequalities will hold:
\begin{equation}
\sum_{i=1}^{N}|x_i|\leq
\sum_{j=1}^{N}\sum_{i=1}^{N}P_{ij}|x_j|<\sum_{j=1}^{N}|x_j|,
\end{equation}
which is a contradiction. 
\section{}
\label{APP:APP_B}
Using an indexing of the nodes that is suitable for the definition of bipartite
graphs, every vector mentioned in the main text can be split into two parts,
$\bp=(\bp_1,\bp_2)^T$, $\bd=(\bd_1,\bd_2)^T$ $\bxi=(\bxi_1,\bxi_2)^T$ such that
the $N_1$ ($N_2$) components of the first (second) part belong to nodes in
$\mathcal{G}_1$ ($\mathcal{G}_2$). The form of $\bA$ and $\bD$ are the
following.
\begin{equation}
\bA =
\begin{pmatrix}
  \bZero & \bB \\
 \bB^T & \bZero
\end{pmatrix}
\quad
\bD =
\begin{pmatrix}
  \bD_1 & \bZero \\
 \bZero & \bD_2
\end{pmatrix},
\end{equation}
where $\bB$ is an $N_1\times N_2$, $\bD_1$ is an $N_1\times N_1$ and $\bD_2$ is
an $N_2\times N_2$ matrix. The transition matrix has the following form:
\begin{equation}
\bA\bD^{-1}  =
\begin{pmatrix}
  \bZero & \bB\bD_2^{-1} \\
 \bB^T\bD_1^{-1} & \bZero
\end{pmatrix}
\end{equation}

The structure of $\bD^{-1/2}\bA\bD^{-1/2}$ is similar to $\bA$, the
off-diagonal block matrices are $\bD_1^{-1/2}\bB\bD_2^{-1/2}$ and its
transpose. A one line calculation shows, that if
$\bu_i=(\bu_{1,k},\bu_{2,k})^T$ is an eigenvector of $\bD^{-1/2}\bA\bD^{-1/2}$
with eigenvalue $\lambda_k$, then the vector $(\bu_{1,k},-\bu_{2,k})^T$, is
also an eigenvector with eigenvalue $-\lambda_k$, so the spectra of
$\bD^{-1/2}\bA\bD^{-1/2}$ is symmetric to the origin. One consequence of this
symmetry is that if the number of nodes is even (odd), the kernel dimension of
$\bD^{-1/2}\bA\bD^{-1/2}$ is also even (odd). For the sake of simplicity, we
study only bipartite graphs with an even number of nodes. For large networks, this
has no serious consequence. Let us define the matrices $\bI^{(1)}_k$ and
$\bI^{(2)}_k$:
\begin{equation}
\bI^{(1)}_k=
\begin{pmatrix}
  \bu_{1,k}\bu_{1,k}^T & \bZero \\
  \bZero & \bu_{2,k}\bu_{2,k}^T
\end{pmatrix},	
\end{equation}
and
\begin{equation}
\bI^{(2)}_k=
\begin{pmatrix}
  \bZero & \bu_{1,k}\bu_{2,k}^T \\
  \bu_{2,k}\bu_{1,k}^T & \bZero
\end{pmatrix}.
\end{equation}
Then, the spectral decomposition of $\bD^{-1/2}\bA\bD^{-1/2}$ is 
\begin{equation}
\sum_{k=1}^{N}\lambda_k (\bI^{(1)}_k+\bI^{(2)}_k)=\sum_{k=1}^{N/2}2\lambda_k \bI^{(2)}_k
\label{EQ:EQ10}
\end{equation}
It is easier to perform the calculation on the spectral decomposition of
$\bD^{-1/2}\bA\bD^{-1/2}$ instead of $\bA\bD^{-1}$ to get an approximation of
$(\bE-\bP)^{-1}$:
\begin{equation}
(\bE-\bP)^{-1}=\bD^{1/2}(\bE-(1-\mu)\bD^{-1/2}\bA\bD^{-1/2})^{-1}\bD^{-1/2}.
\end{equation}
The spectral decomposition of the factor in the middle of the r.~h.~s.~ is
\begin{equation}
\sum_{k=1}^{N}\frac{\bu_k\bu_k^T}{1-(1-\mu)\lambda_i},
\end{equation}
which, using the symmetry of the spectra, can be written in the form 
\begin{equation}
\sum_{k=1}^{N/2}2\frac{\bI^{(1)}_k+(1-\mu)\lambda_k\bI^{(2)}_k}{1-(1-\mu)^2\lambda_k^2}.
\end{equation}
This also can be split into three parts as in \ref{3T}:
\begin{equation}
\begin{aligned}
  \sum_{\lambda_k=0}2\bI^{(1)}_k&+\frac{2}{\mu}\frac{\bI^{(1)}_k+(1-\mu)\bI^{(2)}_k}{2-\mu}+\\
   &+\sum_{\lambda_k\neq 0,1}2\frac{\bI^{(1)}_k+(1-\mu)\lambda_k\bI^{(2)}_k}{1-(1-\mu)^2\lambda_k^2},
\end{aligned}
\end{equation}
but here, summation runs over only the first half of the spectra. The power
series expansion of the summands of the last term is
\begin{equation}
 2\bI^{(1)}_k+2(1-\mu)\lambda_k\bI^{(2)}_k+2(1-\mu)^2\lambda^2_k\bI^{(1)}_k+\dots
\label{PS2}
\end{equation}
Dropping all the terms except the first gives  
\begin{equation}
\bE+\frac{2}{\mu}\frac{(1-\mu)^2\bI^{(1)}_1+(1-\mu)\bI^{(2)}_1}{2-\mu}
\end{equation}
as the inverse of $\bE-(1-\mu)\bD^{-1/2}\bA\bD^{-1/2}$, so
\begin{equation}
(\bE-\bP)^{-1}\simeq\bE+\frac{1}{\mu}\frac{(1-\mu)^2\bJ^{(1)}_1+(1-\mu)\bJ^{(2)}_1}{2-\mu},
\end{equation}
where $\bJ^{(1)}_1$ and $\bJ^{(2)}_1$ are the following matrices:
\begin{equation}
\bJ^{(1)}_1=
\frac{1}{M}
\begin{pmatrix}
  \bd_{1}\bOne_1^T & \bZero \\
  \bZero & \bd_2\bOne_2^T
\end{pmatrix},	
\end{equation}
and
\begin{equation}
\bJ^{(2)}_1=
\frac{1}{M}
\begin{pmatrix}
  \bZero & \bd_1\bOne_2^T \\
  \bd_2\bOne_1^T & \bZero
\end{pmatrix}.
\end{equation}
This gives the values of $\bxi_1$ and $\bxi_2$:
\begin{equation}
\begin{aligned}
\bxi_1&=\bp_1+\frac{1}{\mu}\frac{(1-\mu)^2\overline{p}_1/\overline{d}_1+(1-\mu)\overline{p}_2/\overline{d}_2}{2-\mu}\bd_1,\\
\bxi_2&=\bp_2+\frac{1}{\mu}\frac{(1-\mu)^2\overline{p}_2/\overline{d}_2+(1-\mu)\overline{p}_1/\overline{d}_1}{2-\mu}\bd_2.
\end{aligned}
\end{equation}
Eq.~\ref{EQ:EQ8} gives the final result for homogeneous loading.

To get the first finite size correction, we have to use not only the first, but
also the second term in \ref{PS2}. Using Eq.~\ref{EQ:EQ10}, the correction term
to the inverse of $\bE-(1-\mu)\bD^{-1/2}\bA\bD^{-1/2}$ appears to be
$(1-\mu)\bD^{-1/2}\bA\bD^{-1/2}-2(1-\mu)\bI^{(2)}_1$, so the corrected formula
for the inverse is  
\begin{equation}
\begin{aligned}
\bE &+\frac{2}{\mu}\frac{(1-\mu)^2\bI^{(1)}_1+(1-\mu)^3\bI^{(2)}_1}{2-\mu}+\\
&+(1-\mu)\bD^{-1/2}\bA\bD^{-1/2},
\end{aligned}
\end{equation}
and the inverse of $\bE-\bP$ is 
\begin{equation}
\begin{aligned}
(\bE-\bP)^{-1}\simeq\bE&+\frac{1}{\mu}\frac{(1-\mu)^2\bJ^{(1)}_1+(1-\mu)^3\bJ^{(2)}_1}{2-\mu}\\
&+(1-\mu)\bA\bD^{-1}.
\end{aligned}
\end{equation}
The corrected values of $\bxi_1$ and $\bxi_2$ are
\begin{equation}
\begin{aligned}
\bxi_1\simeq\bp_1&+\frac{1}{\mu}\frac{(1-\mu)^2\overline{p}_1/\overline{d}_1+(1-\mu)^3\overline{p}_2/\overline{d}_2}{2-\mu}\bd_1+\\
&+(1-\mu)\bB\bD_2^{-1}\bOne_1\\
\bxi_2\simeq\bp_2&+\frac{1}{\mu}\frac{(1-\mu)^2\overline{p}_2/\overline{d}_2+(1-\mu)^3\overline{p}_1/\overline{d}_1}{2-\mu}\bd_2+\\
&+(1-\mu)\bB^T\bD_1^{-1}\bOne_2
\end{aligned}
\end{equation}
In the case of homogeneous loading probabilities, this leads to the appearance
of the harmonic means: 
\begin{equation}
\begin{aligned}
\frac{\xi_{1,i}}{p}\simeq 1&+\frac{1}{\mu}\frac{(1-\mu)^2/\overline{d}_1+(1-\mu)^3/\overline{d}_2}{2-\mu}+\\
&+(1-\mu)d_i/h_i,\\
\frac{\xi_{2,i}}{p}\simeq 1&+\frac{1}{\mu}\frac{(1-\mu)^2/\overline{d}_2+(1-\mu)^3/\overline{d}_1}{2-\mu}d_i+\\
&+(1-\mu)d_i/h_i,
\end{aligned}
\end{equation}
and the individual critical loading probabilities in $\mathcal{G}_1$ and
$\mathcal{G}_2$ are those that are in Eq.~\ref{EQ:EQ9}.

\section{}
\label{APP:APP_C}
The following algorithm calculates the order parameter. Suppose that
$\bp=p\be$, where $\be$ is a normalized vector. If $p$ is close to zero, the
network is in the uncongested phase, the components of $\bxi(p)$ satisfy
Eq.~\ref{EQ:EQ2} and the number of the uncongested nodes is equal to the
number of the nodes in the system. If one increases $p$ slowly, one finds that
one or more nodes will surely have at least one waiting particle in their
queues in the stationary regime, \emph{i.~e.}~at least one $\xi_i(p)$ becomes
$1$ at a certain value of $p$. Increasing $p$ toward this value drives these
nodes to the congested state, and the expected value of the growing length at
the queues at these nodes in one time step is $pe_i+\sum_{j}P_{ij}-1$. On the
other hand, these congested nodes send particles to their neighbors with rates
equal to the corresponding element of $\bP$. These ideas suggest the following
algorithm.     
\begin{enumerate}
\item 
  Set $\bP^{(0)}=\bP$, $\by^{(0)}=\be$, $\bz^{(0)}=\bZero$ and $s$ to a small positive number.
\item In the $k$th step, calculate the vector 
  \begin{equation}
  \bx ^{(k)}(s)=(\bE^{(k)}-\bP^{(k)})^{-1}(s\by^{(k)}+\bz^{(k)}).
  \end{equation}
  Starting from the last value at the $(k-1)$th step, increase $s$ until one of
  the components of $\bx^{(k)}$ becomes  $1$, or $s$ becomes $p$. If the latter
  is the case, equate the components of $\bxi$ to the corresponding components of
  $\bx^{(k)}$. If the former is true, set $\xi_i=1$ at the node where $x^{(k)}_i$
  is equal to one - this is the new congested node. Increase every component of
  $\bz^{(k)}$ with the corresponding element of $\bP$ located in the column of
  the new congested node. Delete the rows and columns of the new congested node
  in $\bE^{(k)}$, $\bP^{(k)}$, $\bx^{(k)}$ and $\bz^{(k)}$ . This gives matrices
  $\bE^{(k+1)}$, $\bP^{(k+1)}$ and vectors $\bx^{(k+1)}$ and $\bz^{(k+1)}$ for
  the $(k+1)$th step.  
\item 
  If all components of $\bxi$ are calculated, the order parameter is
  \begin{equation}
  \eta (\bp)=\frac{\sum_{\xi_i=1}\left (p_i+\sum_{j}P_{ij}\xi_j-1 \right )}{\sum_ip_i}.
  \end{equation}
\end{enumerate}

\bibliography{Barankai_Fekete_Vattay-Effect_of_structure}
\bibliographystyle{apsrev4-1}
\end{document}